\documentclass[twoside]{article}
\usepackage{graphicx,bm,color,latexsym,amssymb}
\usepackage{qic,epsfig}
\usepackage[latin1]{inputenc}
\usepackage{subfigure}

\definecolor{Blue}{rgb}{0.0,0.0,1}
\definecolor{Red}{rgb}{1,0.0,0.0}
\definecolor{Green}{rgb}{0,0.5,0.0}
\definecolor{Yellow}{rgb}{0.5,0.5,0.0}

\usepackage[matrix,frame,arrow]{xy}
\usepackage{amsmath}

\newcommand{\qw}[1][-1]{\ar @{-} [0,#1]}
\newcommand{\qwx}[1][-1]{\ar @{-} [#1,0]}

\newcommand{\gate}[1]{*{\xy *+<.6em>{#1};p\save+LU;+RU **\dir{-}\restore\save+RU;+RD **\dir{-}\restore\save+RD;+LD **\dir{-}\restore\POS+LD;+LU **\dir{-}\endxy} \qw}

\newcommand{\control}{*-=-{\bullet}}
\newcommand{\controlo}{*!<0em,.04em>-<.07em,.11em>{\xy *=<.45em>[o][F]{}\endxy}}
\newcommand{\ctrl}[1]{\control \qwx[#1] \qw}
\newcommand{\ctrlo}[1]{\controlo \qwx[#1] \qw}
\newcommand{\targ}{*{\xy{<0em,0em>*{} \ar @{ - } +<.4em,0em> \ar @{ - } -<.4em,0em> \ar @{ - } +<0em,.4em> \ar @{ - } -<0em,.4em>},*+<.8em>\frm{o}\endxy} \qw}
\newcommand{\qswap}{*=<0em>{\times} \qw}
\newcommand{\multigate}[2]{*+<1em,1.4em>{\hphantom{#2}} \qw \POS[0,0].[#1,0];p !C *{#2},p \save+LU;+RU **\dir{-}\restore\save+RU;+RD **\dir{-}\restore\save+RD;+LD **\dir{-}\restore\save+LD;+LU **\dir{-}\restore}
\newcommand{\ghost}[1]{*+<1em,1.4em>{\hphantom{#1}} \qw}
\newcommand{\Qcircuit}{\xymatrix @*=<0em>}

\begin{document}
\setlength{\textheight}{8.0truein}    

\runninghead{Title  $\ldots$}
            {Author(s) $\ldots$}

\normalsize\textlineskip
\thispagestyle{empty}
\setcounter{page}{1}


\vspace*{0.88truein}


\fpage{1}

\centerline{\bf NUCLEAR SPIN $3/2$ ELECTRIC QUADRUPOLE  RELAXATION }
\vspace*{0.035truein}
\centerline{\bf AS A QUANTUM COMPUTATION}
\vspace*{0.37truein}
\centerline{\footnotesize A. M. Souza, A. Gavini-Viana, I. S. Oliveira and R. S. Sarthour}
\vspace*{0.015truein}
\centerline{\footnotesize\it  Centro Brasileiro de
Pesquisas F\'\i sicas, Rua Dr. Xavier Sigaud 150 }
\baselineskip=10pt
\centerline{\footnotesize\it   Rio de Janeiro 22290-180, RJ, Brazil }
\vspace*{10pt}
\centerline{\footnotesize 
R. Auccaise, J. Teles, E. R. deAzevedo and T. J. Bonagamba}
\vspace*{0.015truein}
\centerline{\footnotesize\it  Instituto de  F\'\i sica de S\~ao Carlos, Universidade de S\~ao Paulo,
Caixa Postal 369}
\baselineskip=10pt
\centerline{\footnotesize\it S\~ao Carlos, 13560-970 SP, Brazil.  }
\vspace*{10pt}

\vspace*{0.225truein}

\vspace*{0.21truein}
\abstracts{
In this work we applied a quantum circuit treatment to describe the 
nuclear spin relaxation. From the Redfield theory, we were able 
to describe the quadrupolar relaxation as a computational process in 
the case of spin 3/2 systems, through a model in which the environment is
comprised by five qubits and three different quantum noise channels. The 
interaction between the
environment and the spin 3/2 nuclei is then described by a quantum circuit
fully compatible with the Redfield theory of relaxation. Theoretical predictions 
are compared to  experimental data, a short review of quantum channels 
and relaxation in NMR qubits is also present.}{}{}

\vspace*{10pt}
\vspace*{3pt}

\vspace*{1pt}\textlineskip	
\section{\label{int} Introduction }
\vspace*{-0.5pt}
\noindent
 The analysis of physical
systems in terms of information processing \cite{loyd1,loyd2} can be very productive and
complementary to the conventional description of nature in terms of
forces and energy. A good example is the famous
Maxwell's demon paradox, only solved using arguments from
information theory \cite{maruyama}. Recent works have applied this
approach to gain novel insights into cosmology and fundamental physics
\cite{loyd1,loyd2,loyd3,hsu,maruyama}. In biophysics, Engel et al. \cite{engel} 
found evidence that photosynthetic plants employ a kind of quantum search algorithm to efficiently 
capture the energy of the Sun. Furthermore, the view of physical phenomena  as quantum 
computational processes can be  
sometimes more adequate to design algorithms to simulate physical systems. One open problem 
in quantum science is the the problem of simulating open systems on quantum computers, considering 
relaxation phenomena in the context of information processing could yield some insight into 
this problem.

Nuclear Magnetic Resonance (NMR) is a well
established technique used in Physics, Chemistry, Medicine, and
Biology. In the last decade, it has also been used as an
experimental method for many Quantum Information Processing (QIP)
implementations \cite{oliveira,suter} and in the study of
fundamental aspects of quantum mechanics \cite{souza,nelson}. While $N$ coupled 
spin $1/2$ nuclei, either in solid or liquid
state, have been extensively used to process information 
of $N$ qubit systems, quadrupolar nuclei
with spin $I > 1/2$, which account for about three quarters of naturally occurring magnetic nuclei, have
also been used to process information of $\log_2(2I +1)$
equivalent qubit systems in liquid 
crystals \cite{jmr,pra,Khitrin,Murali,Khitrin2,Gopinath}, solids \cite{Kampermann} and more 
recently in GaAs nanodevices \cite{yusa,Leuenberger,Kondo,Hirayama}.

Here we have used the quantum information view to describe the 
relaxation of NMR qubits. In this work, we describe a NMR relaxation process using the
``language" of quantum circuits, following the approach of quantum
information processing. For this study, the relaxation of a
quadrupolar nucleus, originated by local electric field fluctuations,
was fully characterized by means of basic quantum logic gates.
The gates are related to some relaxation parameters, such as
the spectral densities and the quadrupolar coupling strength, which contain all the information about the
relaxation process. In the present paper we demostrate that the relaxation of spin 3/2 nuclei can be described by a model in which the environment is
comprised by five qubits and three different quantum noise channels. The 
interaction between the environment and the nuclei is then described by a quantum circuit
fully compatible with the Redfield theory of relaxation.

This paper is organized as follows: In sections \ref{qchan} and
\ref{qnuc}, we present a short review of quantum channels and a
brief description of the nuclear spin quadrupole 
relaxation. Secs. \ref{qmodel} and \ref{expd} contain the
description of the quadrupolar relaxation as a computational process
and a comparison between the theoretical model and experimental
results. Finally, in the last section, some conclusion are drawn.

\section{\label{qchan} Quantum Channels}
\noindent
The dynamics of a quantum system ($\mathcal{S}$) that interacts
with an environment ($\mathcal{E}$) can be described by
the following Hamiltonian:

\noindent 
\begin{equation}
\mathcal{H} = \mathcal{H}_s +  \mathcal{H}_e + \mathcal{H}_{int},
\end{equation}
where $\mathcal{H}_s$ and $\mathcal{H}_e$ are the system and
environment internal Hamiltonians, respectively, and
$\mathcal{H}_{int}$ is the coupling Hamiltonian between them. The
set system-environment is a closed system and must evolve unitarily
in time, according to the laws of quantum mechanics. This process
can be represented using a quantum circuit, as may be seen in Figure
\ref{circm}. The unitary evolution $U = e^{-i\mathcal{H}t/\hbar }$
can be interpreted as a quantum computational process and therefore be broken
into smaller quantum logical gates. Due to the action of $U$, after
some time $t$, both $\mathcal{S}$ and $\mathcal{E}$ become entangled
and an initially pure state of $\mathcal{S}$  can  be turned into a
mixed state \cite{nielsen}.

In real situations, the actual form of $U$, i.e. the quantum algorithm
computed, depends on the specific processes that can take place between 
$\mathcal{S}$ and $\mathcal{E}$.  Although the environment
degrees of freedom are usually very large, it is possible to model
the environment with a finite number of qubits \cite{nielsen}. In
fact,  at most $d^2$ qubits are necessary to model the environment,
where $d$ is the Hilbert space dimension of $\mathcal{S}$. 

Denoting $\rho$ and $\rho_{env}$ as the initial states of $\mathcal{S}$ and $\mathcal{E}$,
respectively, one can show, by tracing over the degrees of freedom of the environment,
that the effective evolution of $\rho$  is not unitary, and is given by:


\begin{figure} [htbp]
\vspace*{13pt}

\begin{center}

\begin{tabular}{ c }
                                   
\Qcircuit @C=2.3em @R=1.7em @!R {  \rho       & & \multigate{1}{U} & \qw  & \sum_k E_k \rho E_{k}^{\dag} \\ 
                                   \rho_{env} & & \ghost{U}   & \qw  & \\ }
\end{tabular}

\end{center}
\vspace*{13pt}
\fcaption{\label{circm} Circuit model for the system-environment
interaction. $\rho$ and $\rho_{env}$ represent, respectively, the initial 
state of the system and the environment.}
\end{figure}


\begin{equation} \label{ek}
\rho' = \sum_k E_k \rho E_k^{\dag},
\end{equation}
where the set of operators $E_k$ are the so-called Krauss operators
and the condition $\sum_k E_k \rho E_k^{\dag} = I$ must be satisfied
in order to preserve the trace of the density matrix. The interpretation of this 
expression is that $\rho$ is transformed into
$E_k \rho E_k^{\dag}$ with probability $Tr(E_k \rho E_k^{\dag})$.
The expression (\ref{ek}) is only valid if the set
system-environment is initially in a non-entangled state. Usually, it
is not the case since the constant interaction between them always
produces quantum correlations. However,  these correlations are
destroyed upon the preparation of the initial state.

Regarding the main system $\mathcal{S}$ as being composed by $N$
subsystems, one can recognize two types of processes: The first,
called Global Channel, is a process in which all the subsystems interact
with the same environment. In this case, the operation $U$ is a
non-separable matrix, the channel can create entanglement between $\mathcal{S}$
and $\mathcal{E}$  and, in principle, either can destroy or create
entanglement among the subsystems. In
contrast, Local Channels are those processes where each subsystem
interacts with its own environment. In this case, entanglement among
subsystems cannot be created since $U = U_1 \otimes U_2 \otimes
\cdots \otimes U_N$ and the Eq. (\ref{ek}) has the form:
\noindent 
\begin{equation} \label{ek2}
\rho = \sum_{k \cdots m} E_{k}^1 \otimes \cdots  \otimes E_{m}^N
\rho E_{k}^{1\dag} \otimes \cdots  \otimes E_{m}^{N\dag}.
\end{equation}

There are many types of quantum channels, for example
\cite{nielsen}, the generalized amplitude damping (GAD) channel, the
phase damping (PD) channel, the bit-flip, phase-flip and
depolarizing channels. We will briefly review now two 
channels - GAD and PD - which are useful to describe NMR relaxation.

\subsection{Generalized Amplitude Damping \label{gadsec}}
\noindent
 The generalized amplitude damping channel describes
 dissipative interactions between the system and its environment
 at finite temperatures \cite{nielsen}. It can be decomposed
 into two processes. The first, denoted here as $\mathcal{A}_{1 \rightarrow 0}$,
 is a process which a qubit in the excited state $|1\rangle$ decays to its
 fundamental state $|0\rangle$ with some probability $\gamma$. In the second
 process ($\mathcal{A}_{0 \rightarrow  1}$), the qubit is excited from the ground
 state with probability $(1 - \gamma)$. The former process is
 called amplitude damping and describes a qubit in contact with a reservoir at
 temperature $T = 0$ K. To introduce finite temperatures, one have to consider that
 the process $\mathcal{A}_{0 \rightarrow  1}$ occurs with probability $\mathcal{P}$ and
 $\mathcal{A}_{1 \rightarrow  0}$ occurs with probability $1-\mathcal{P}$,
 where $\mathcal{P}$ is the probability of finding the system, at thermal equilibrium, in
 its ground state.

 The Krauss operators for the GAD channel are described
in Eq. (\ref{gadk1})-(\ref{gadk4}) \cite{nielsen}. The circuit models for
the processes $\mathcal{A}_{0 \rightarrow  1}$ and $\mathcal{A}_{1 \rightarrow  0}$ are
 shown in Figures (\ref{gad1}) and (\ref{gad2}). The circuit model for GAD can be constructed
 combining both processes (see Figure (\ref{gad3})).

\noindent 
\begin{eqnarray}
E_1 &=& \sqrt{\mathcal{P}} \left ( \begin{array}{cc} 1 &0 \\ 0 & \sqrt{1-\gamma} \end{array}\right), \label{gadk1} \\
E_2 &=& \sqrt{\mathcal{P}} \left ( \begin{array}{cc} 0 &\sqrt{\gamma} \\ 0 & 0 \end{array}\right),  \label{gadk2} \\
E_3 &=& \sqrt{1-\mathcal{P}} \left ( \begin{array}{cc} \sqrt{1-\gamma} &0 \\ 0 & 1 \end{array}\right), \label{gadk3} \\
E_4 &=& \sqrt{1-\mathcal{P}} \left ( \begin{array}{cc} 0 &0 \\  \sqrt{\gamma} & 0\end{array}\right).
\label{gadk4}
\end{eqnarray}


\begin{figure}[htbp]
\vspace*{13pt}
\begin{center}

\begin{tabular}{c}
{\Qcircuit @C=2.3em @R=1.7em @!R { \rho &  & \ctrl{1}           & \targ &  \qw   \\        
                          | 0 \rangle   &  & \gate{R_y(\alpha)} &  \ctrl{-1} & \qw  \\} }
\end{tabular}

\end{center}
\vspace*{13pt}
\fcaption{\label{gad1} Quantum circuit description of the
process  $\mathcal{A}_{1 \rightarrow  0}$. The initial state of
the system is represented by $\rho$ while the environment is initialized in
the pure state $|0\rangle$. The notation $R_y(\alpha)$ represents
a $\alpha = 2\arcsin(\sqrt{\gamma})$ rotation around the axis $y$.}
\end{figure}


\begin{figure}[htbp]
\vspace*{13pt}
\begin{center}
\begin{tabular}{c}
{\Qcircuit @C=2.3em @R=1.7em @!R { \rho & &\qw      & \ctrlo{1}         & \targ          & \gate{Z}& \qswap & \qw &  \\        
                          | 0 \rangle   & &\gate{X} & \gate{R_y(\beta)}  &  \ctrlo{-1} &  \ctrl{-1}& \qswap \qwx & \qw  \\} }
\end{tabular}
\end{center}
\vspace*{13pt}
\fcaption{\label{gad2} Quantum circuit description of the
process  $\mathcal{A}_{0 \rightarrow  1}$. The initial state of
the system is represented by $\rho$ while the environment is initialized in
the pure state $|0\rangle$. The notation $R_y(\beta)$ represents
a $\beta = 2\arcsin(\sqrt{\gamma}) +\pi$ rotation around the axis $y$.}
\end{figure}

\begin{figure*}[htbp]
\vspace*{13pt}
\begin{center}
\begin{tabular}{ c }
\Qcircuit @C=1.3em @R=1.0em @!R { \rho  & &  \ghost{\mathcal{A}_{0 \rightarrow 1 }} & \ghost{\mathcal{A}_{1 \rightarrow 0 }} & \qw   &  & & \ctrl{1}          &\targ     & \qw   &\ctrlo{1}&\targ &\gate{Z}&\qswap & \qw \\   
                                       | 0 \rangle & & \multigate{-1}{\mathcal{A}_{0  \rightarrow 1}} & \multigate{-1}{\mathcal{A}_{1  \rightarrow  0 }} & \qw & =& & \gate{R_y(\alpha)}&\ctrl{-1} & \targ      &\gate{R_y(\beta)}& \ctrlo{-1} &\ctrl{-1}& \qswap \qwx & \qw \\
 |\Phi \rangle_{eq} & &  \ctrlo{-1} &  \ctrl{-1} & \qw &  & & \ctrlo{-1}       &\ctrlo{-1} &\ctrl{-1} &\ctrl{-1} & \ctrl{-1}&\ctrl{-1} & \ctrl{-1} & \qw\\  }
\end{tabular}
\end{center}
\vspace*{13pt}
\fcaption{\label{gad3} Quantum circuit description of the generalized amplitude
damping channel. The initial state of the system is represented by $\rho$ while
the environment is initialized in the 
pure state $| 0 \rangle \otimes | \Phi \rangle_{eq}$, where $| \Phi \rangle_{eq}  = \sqrt{\mathcal{P}}| 0 \rangle + \sqrt{1-\mathcal{P}}| 1 \rangle $.}
\end{figure*}

\subsection{Phase Damping \label{pdsec} }
\noindent
The phase damping channel describes the loss of coherence without
loss of energy \cite{nielsen}. In this channel, the relative phase
between $|0 \rangle$ and $|1 \rangle$  remains unchanged with some
probability $\lambda$ or is inverted ($\phi \rightarrow  \phi +
\pi$) with  probability  $1 - \lambda$. The states of the
computational basis do not change under this process. However,
superpositions in the computational basis  can get entangled with the environment. Thus,
this channel does not change the probability of finding the qubit in
the state $|0\rangle$ or $|1\rangle$, but it destroys all coherences
between them. The quantum circuit model for this channel is shown in
Figure (\ref{pd1}) and their Krauss operators are given by
\cite{nielsen}:

\noindent 
\begin{eqnarray}
\label{pdk}
E_{1}  &=&  \sqrt{\lambda }\left( \begin{array}{cc} 1 & 0\\ 0 & 1 \end{array} \right), \\
E_{2}  &=&  \sqrt{1-\lambda } \left( \begin{array}{cc} 1 & 0  \\ 0 & -1 \end{array} \right).
\end{eqnarray}

\begin{figure}[htbp]
\vspace*{13pt}
\begin{center}

\begin{tabular}{c}
\Qcircuit @C=1.7em @R=1.2em@!R { \rho       & & \ctrl{1}    & \qw  \\        
                                 | 0 \rangle & & \gate{R_y(\theta)} & \qw \\}
\end{tabular}

\end{center}
\vspace*{13pt}
\fcaption{\label{pd1} Quantum circuit description of the  phase damping channel. The initial
state of the system is represented by $\rho$ while the environment is initialized in
the pure state $|0\rangle$. The notation $R_y(\theta)$ represents
a $\theta = 2\arccos(2\lambda - 1)$ rotation around the axis $y$.}
\end{figure}

 \subsection{Bloch Equation and Quantum Channels \label{blochsec}}
\noindent
In general, two different processes occur simultaneously
during the relaxation in NMR systems: the transverse relaxation and the
longitudinal relaxation. The first leads the disappearance of the
 nuclear magnetization components ($M_x$ and $M_y$) that are
perpendicular to the strong static field applied along 
the $z$ direction. This process causes
decoherence without energy exchange with the environment. The
longitudinal relaxation leads the ensemble of nuclear spins to
return to its equilibrium state. This process is related to
transitions between the nuclear Zeeman energy levels. Unlike 
the transverse relaxation, the longitudinal relaxation is a
mechanism where the system exchanges energy with the environment. In
order to show an example of how the longitudinal and transverse
relaxations can be related to the quantum channels, let's consider the
relaxation of a single nuclear spin $1/2$ in the initial state

\noindent
\begin{eqnarray}
\label{mat1}
\left ( 
\begin{array}{ccc}
M_z^{0} & M_x^{0} + iM_y^{0} \\ M_x^{0} - iM_y^{0} & 1-M_z^{0} 
\end{array}
 \right ),
\end{eqnarray}
where $M_j^{0}$ is the initial magnetization along $j=x,y,z$ axis. The application of 
GAD and PD channels on (\ref{mat1}) leads to:
\noindent
\begin{eqnarray}
\label{mag1}
M_x  &=&  M_x^{0} \sqrt{ 1-\gamma}( 2\lambda-1)  ,\\
M_y  &=&  M_y^{0} \sqrt{ 1-\gamma}( 2\lambda-1)  , \label{mag2} \\
M_z  &=&  M_z^{0} (1-\gamma) + \gamma( 2 \mathcal{P} -1). \label{mag3}
\end{eqnarray}

One can show that (\ref{mag1})-(\ref{mag3}) reproduce the solution of
phenomenological Bloch equations \cite{oliveira} if $\gamma = 1-e^{-t/T_1}$,
$\lambda = (1+e^{-t/\alpha})/2 $ and $\alpha = 2T_1 T_2/(2T_1-T_2$), where
$T_1$ and $T_2$ are, respectively, the longitudinal
and transverse relaxation times \cite{oliveira}. Consequently, one can
combine the circuits (\ref{gad3}) and (\ref{pd1}) to
design a quantum circuit that describes the relaxation of a single spin
$1/2$ and is completely equivalent to the Bloch equations
description. The circuit we have derived is shown in Figure (\ref{bloch}).

\begin{figure*}[htbp]
\vspace*{13pt}
\begin{center}

\begin{tabular}{ c }
\Qcircuit @C=1.7em @R=1.2em @!R { \rho  & & \ctrl{1} & \ctrl{2}          &\targ     & \qw    &\ctrlo{2}& \targ &\gate{Z}&\qswap & \qw \\   
                            | 0 \rangle & & \gate{R_y(\theta)} & \qw & \qw & \qw & \qw & \qw & \qw & \qw \qwx & \qw \\
                            | 0 \rangle & & \qw & \gate{R_y(\alpha)}&\ctrl{-2} & \targ      &\gate{R_y(\beta)}& \ctrlo{-2} & \ctrl{-2}& \qswap \qwx & \qw \\
                            |\Phi \rangle_{eq} & & \qw &  \ctrlo{-1}       &\ctrlo{-1} &\ctrl{-1} &\ctrl{-1} & \ctrl{-1}& \ctrl{-1} & \ctrl{-2} & \qw\\  }
\end{tabular}

\end{center}
\vspace*{13pt}
\fcaption{\label{bloch} A quantum circuit model equivalent to the Bloch
equations description of relaxation for a single spin $1/2$. The circuit can be
constructed combining figures (\ref{gad3}) and (\ref{pd1}). }
\end{figure*}

\section{Nuclear Electric Quadrupole Relaxation \label{qnuc}}
\noindent
Nuclei with spin $I > 1/2$  have an asymmetric charge distribution and thus
possess an electric quadrupole moment $Q$. Therefore, a quadrupolar nucleus can
interact with either magnetic fields or electric field gradients \cite{abragam,Gerothanassis}. Usually in 
liquid state, rapid molecular tumbling tends to completely average the
electric field gradients and consequently, this interaction
is not observed in the spectrum. However, even when any quadrupolar 
splintting is observed, the relaxation due to electric field gradient 
fluctuations can significantly contribute to the longitudinal 
and transverse relaxation. In many cases of interest, the averaging is not complete 
and an residual axially symmetric electric field gradient can be 
effective. In this cases, the dynamics of a
quadrupolar nucleus is described by the Hamiltonian \cite{oliveira}:

\noindent
\begin{equation}  \label{Hq}
\mathcal{H}=-\hbar \omega _{L} \mathcal{I}_{z}   + \frac{\hbar \omega _{Q}}{6}\left( 3\mathcal{I}%
_{z}^{2}-\mathcal{I}^{2}\right),
\end{equation}
where $\omega_L$ and $\omega_Q$ are, respectively, the Larmor and
quadrupole frequencies. For a spin 3/2 system, such as the $^{23}$Na nuclei, this Hamiltonian
gives rise to four unequally spaced energy levels, originating an 
NMR spectrum containing three
lines, corresponding to transitions between adjacent levels. The
energy states $\left|3/2\right\rangle$, $\left|1/2\right\rangle$,
$\left|-1/2\right\rangle$, and $\left|-3/2\right\rangle$, can be
labeled as $\left|00\right\rangle$, $\left|01\right\rangle$,
$\left|10\right\rangle$, and $\left|11\right\rangle$, which
correspond to a two-qubit system.

Nuclear relaxation is caused by the interaction between the nuclear spin
and random electromagnetic fields generated by the
environment. The relaxation can occur through several mechanisms, depending on
the type of interactions existing in the spin system \cite{abragam}. When the quadrupolar coupling
strength is much higher than the dipolar fields, the loss of
coherence and energy dissipation  can then be considered 
to be exclusively due to electric field gradient fluctuations
\cite{abragam,Gerothanassis,roberto}. Under
pure quadrupolar relaxation mechanism, one can show, using the Redfield
formalism (\cite{abragam,redfield}), that the time evolution of 
each element of a spin $3/2$ density matrix is given by \cite{ruben} the equations (\ref{Rho01})-(\ref{Rho33}). The
superscript $eq$ denotes the thermal equilibrium state and the index
values $i,j=0,1,2,3$ corresponds to the states
$\left|3/2\right\rangle$, $\left|1/2\right\rangle$,
$\left|-1/2\right\rangle$, and $\left|-3/2\right\rangle$,
respectively.

\noindent
\begin{eqnarray}
 {\rho } _{01}\left( t \right) = \frac{1}{2} [ {\rho }_{01}\left( t_{0}\right)+{\rho } _{23}\left( t_{0}\right)  \nonumber \\
 +\left( {\rho}_{01}\left( t_{0}\right)-{\rho } _{23}\left( t_{0}\right)\right)e^{-2CJ_{2}\left(t-t_{0}\right)} ]e^{-C\left( J_{0}+J_{1}\right) \left( t-t_{0}\right)} , \label{Rho01} \\
{\rho } _{23}\left( t\right) =\frac{1}{2} [ {\rho }_{01}\left( t_{0}\right)+{\rho } _{23}\left( t_{0}\right) \nonumber \\
                              -\left( {\rho}_{01}\left( t_{0}\right)-{\rho } _{23}\left( t_{0}\right)\right)
e^{-2CJ_{2}\left(t-t_{0}\right)} ]e^{-C\left( J_{0}+J_{1}\right) \left( t-t_{0}\right)} , \label{Rho23} \\
{\rho } _{02}\left( t\right) =\frac{1}{2}[ {\rho }_{02}\left( t_{0}\right)+{\rho } _{13}\left( t_{0}\right) \nonumber \\
                              +\left( {\rho}_{02}\left( t_{0}\right)-{\rho } _{13}\left( t_{0}\right)\right) e^{-2CJ_{1}\left( t-t_{0}\right)} ]e^{-C\left( J_{0}+J_{2}\right) \left( t-t_{0}\right)} , \label{Rho02} \\
{\rho } _{13}\left( t\right) =\frac{1}{2}[ {\rho }_{02}\left( t_{0}\right)+{\rho } _{13}\left( t_{0}\right) \nonumber \\
                              -\left( {\rho}_{02}\left( t_{0}\right)-{\rho } _{13}\left( t_{0}\right)\right) e^{-2CJ_{1}\left( t-t_{0}\right)} ]e^{-C\left( J_{0}+J_{2}\right) \left( t-t_{0}\right)} , \label{Rho13} \\
{\rho } _{12}\left( t\right) ={\rho } _{12}\left(t_{0}\right)e^{-C\left( J_{1}+J_{2}\right) \left( t-t_{0}\right)}
,\label{Rho12}\\
{\rho } _{03}\left( t\right) ={\rho } _{03}\left(t_{0}\right)e^{-C\left( J_{1}+J_{2}\right) \left( t-t_{0}\right)}
,\label{Rho03} \\
{\rho }_{00}\left( t\right)  = \rho_{00}^{eq} - \frac{1}{4} [ R_{1}^{0}e^{-2C\left( J_{1}+J_{2}\right)\left( t-t_{0}\right)} \nonumber \\
                              -R_{2}^{0}e^{-2CJ_{2}\left( t-t_{0}\right)}-R_{3}^{0}e^{-2CJ_{1}\left(t-t_{0}\right) } ] , \label{Rho00} \\
{\rho }_{11}\left( t\right)  =   \rho_{11}^{eq} +\frac{1}{4} [ R_{1}^{0}e^{-2C\left( J_{1}+J_{2}\right) \left( t-t_{0}\right)} \nonumber \\
                              +R_{2}^{0}e^{-2CJ_{2}\left( t-t_{0}\right)}-R_{3}^{0}e^{-2CJ_{1}\left(t-t_{0}\right) } ] , \label{Rho11} \\
 {\rho }_{22}\left( t\right)  =   \rho_{22}^{eq} +\frac{1}{4} [R_{1}^{0}e^{-2C\left( J_{1}+J_{2}\right) \left(t-t_{0}\right) } \nonumber \\
                               -R_{2}^{0}e^{-2CJ_{2}\left( t-t_{0}\right)}+R_{3}^{0}e^{-2CJ_{1}\left(
t-t_{0}\right) } ] , \label{Rho22} \\
 {\rho }_{33}\left( t\right)  =   \rho_{33}^{eq} -\frac{1}{4} [R_{1}^{0}e^{-2C\left( J_{1}+J_{2}\right) \left(t-t_{0}\right) } \nonumber \\
                               +R_{2}^{0}e^{-2CJ_{2}\left( t-t_{0}\right)}+R_{3}^{0}e^{-2CJ_{1}\left( t-t_{0}\right) } ] .
\label{Rho33}
\end{eqnarray}
where
\noindent 
\begin{eqnarray}
R_1^0  &=& -(\rho_{00} - \rho_{00}^{eq}) + (\rho_{11} - \rho_{11}^{eq}) + (\rho_{22} - \rho_{22}^{eq}) \nonumber \\
& & - (\rho_{33} - \rho_{33}^{eq}) \\
R_2^0  &=& (\rho_{00} - \rho_{00}^{eq}) + (\rho_{11} - \rho_{11}^{eq}) - (\rho_{22} - \rho_{22}^{eq})\nonumber  \\
& & - (\rho_{33} - \rho_{33}^{eq}) \\
R_3^0  &=& (\rho_{00} - \rho_{00}^{eq}) - (\rho_{11} - \rho_{11}^{eq}) + (\rho_{22} - \rho_{22}^{eq}) \nonumber \\
& & - (\rho_{33} - \rho_{33}^{eq})
 \label{RRR}
\end{eqnarray}

The parameters $J_0$, $J_1$ and $J_2$ are the spectral densities \cite{maarel}
and  $C$ is a proportionality coefficient that can be determined
using the quadrupolar coupling strength \cite{jaccard}.

\section{Quantum Computing Model for the Quadrupolar Relaxation \label{qmodel}}
\noindent
In the work of Auccaise et al. \cite{ruben}, it was shown that
the longitudinal relaxation of each qubit in a spin $3/2$
nucleus under pure quadrupolar relaxation are related to the 
spectral densities $J_2$ and $J_1$, respectively, and occurs
independently. These features suggest that the longitudinal relaxation 
in this case can be described by two dissipative 
GAD (see section \ref{gadsec}) channels acting on each qubit separately.

In contrast, the loss of coherence depends 
on the three spectral densities. By
inspection of Equations (\ref{Rho01})-(\ref{Rho03}), one can see
that only the coherence elements $\rho_{01}$,
$\rho_{02}$ , $\rho_{13}$ and  $\rho_{23}$ depend on $J_0$, suggesting 
that the phase damping channel in this case does not  destroy the coherence between the states $|00 \rangle$ and 
$|11 \rangle$ and between $|01 \rangle$ and 
$|10 \rangle$, in other words we can say that the Bell's states form a free decoherence subspace 
for such channel. A possible quantum channel to reproduce this feature is a channel
where the relative phase between $|0 \rangle$ and $|1 \rangle$ in both qubits remains unchanged with
some probability $\lambda$ or are simultaneously inverted with  probability  $1 - \lambda$. We will call this
channel as  Global Phasing Damping (GPD). The
quantum circuit model for the GPD channel is presented in Figure
(\ref{gpd}) and its Krauss operators are given by:

\noindent
\begin{eqnarray}
\label{gpdop}
E_{0} &=&\sqrt{1-\lambda }\left(
\begin{array}{cccc}
1 & 0 & 0 & 0\\
0 & -1& 0 & 0\\
0 & 0 &-1 & 0\\
0 & 0 & 0 & 1
\end{array}
\right) \\
E_{1} &=\sqrt{\lambda }&\left(
\begin{array}{cccc}
1 & 0 & 0 & 0\\
0 & 1 & 0 & 0\\
0 & 0 & 1 & 0\\
0 & 0 & 0 & 1
\end{array}
\right)
\end{eqnarray}

Therefore, the quadrupolar relaxation process can be described by a model with 
three different channels:
two local channels related to dissipation of energy and one global
channel related to loss of coherence without energy exchange. From the above statement, we can 
conclude that the quadrupolar relaxation can lead to sudden death of entanglement. Since the Bell states 
are not affected by the phase damping channel, the loss of coherence in such states occurs due to  
two independent GAD channels. As demonstrated in \cite{Asma}, under this situation all states undergo 
sudden death of entanglement. One 
can show, using Eq. (\ref{ek}) and (\ref{ek2}), that under the
action of these channels, the evolution 
of each density matrix element is given by:

\noindent 
\begin{eqnarray}
\rho_{01}(t) &=& [ ( 1 - \gamma_A (1-\mathcal{P}_A) ) \rho_{01}(t_0) \nonumber \\ &&+ (\gamma_{A} \mathcal{P}_A) \rho_{23}(t_0)  ] \sqrt{1-\gamma_B}(2 \lambda -1), \label{eksol1}\\
\nonumber  \\
\rho_{23}(t) &=& [ ( \gamma_A (1-\mathcal{P}_A) ) \rho_{01}(t_0) \nonumber \\ &&+ (1 - \gamma_{A} \mathcal{P}_A) \rho_{23}(t_0)  ] \sqrt{1-\gamma_B}(2 \lambda -1),\label{eksol2} \\
\nonumber  \\
\rho_{02}(t) &=& [ ( 1 - \gamma_B (1-\mathcal{P}_B)  ) \rho_{02}(t_0) \nonumber \\&& + (\gamma_{B}  \mathcal{P}_B )\rho_{13}(t_0)  ] \sqrt{1-\gamma_A}(2 \lambda -1), \label{eksol3}\\
\nonumber \\\
\rho_{13}(t) &=& [ \gamma_{B} (1 - \mathcal{P}_B) \rho_{02}(t_0) \nonumber \\ && +  ( 1 - \gamma_B \mathcal{P}_B  ) \rho_{13}(t_0) \nonumber  ] \sqrt{1-\gamma_A}(2 \lambda -1), \label{eksol4} \\
\rho_{12}(t) &=& \rho_{12}(t_0)\sqrt{1-\gamma_A}\sqrt{1-\gamma_B}, \label{eksol5} \\
\nonumber \\
\rho_{03}(t) &=& \rho_{03}(t_0)\sqrt{1-\gamma_A}\sqrt{1-\gamma_B}, \label{eksol6}\\
\nonumber  \\
\rho_{00}(t) &=&(1-\gamma_A (1-\mathcal{P}_A) ) (1-\gamma_B (1-\mathcal{P}_B) )\rho_{00}(t_0) +\nonumber\\             & &(1-\gamma_A (1-\mathcal{P}_A) ) (\gamma_B \mathcal{P}_B ) \rho_{11}(t_0) +\nonumber \\
             & &(\gamma_A \mathcal{P}_A ) (1-\gamma_B (1-\mathcal{P}_B) )\rho_{22}(t_0) + \nonumber \\
             & &(\gamma_A \mathcal{P}_A) (\gamma_B \mathcal{P}_B ) \rho_{33}(t_0),  \label{eksol7}  \\ \nonumber\\
\rho_{11}(t) &=&(1-\gamma_A (1-\mathcal{P}_A) ) (\gamma_B (1-\mathcal{P}_B) )\rho_{00}(t_0) +\nonumber \\
             & &(1-\gamma_A (1-\mathcal{P}_A) ) (1- \gamma_B \mathcal{P}_B ) \rho_{11}(t_0) +\nonumber \\
             & &(\gamma_A \mathcal{P}_A ) (\gamma_B (1-\mathcal{P}_B) ) \rho_{22}(t_0) + \nonumber \\
             & &(\gamma_A \mathcal{P}_A) (1 - \gamma_B \mathcal{P}_B ) \rho_{33}(t_0),  \label{eksol8}  \\ \nonumber \\
\rho_{22}(t) &=&(\gamma_A (1-\mathcal{P}_A) ) (1-\gamma_B (1-\mathcal{P}_B) )\rho_{00}(t_0) +\nonumber \\
             & &(\gamma_A (1-\mathcal{P}_A) ) (\gamma_B \mathcal{P}_B ) \rho_{11}(t_0) +\nonumber \\
             & &(1-\gamma_A \mathcal{P}_A) ) (1-\gamma_B(1 - \mathcal{P}_B) ) \rho_{22}(t_0) + \nonumber \\
             & &(1-\gamma_A \mathcal{P}_A) (\gamma_B \mathcal{P}_B ) \rho_{33}(t_0),  \label{eksol9}  \\ \nonumber \\
\rho_{33}(t) &=&(\gamma_A (1-\mathcal{P}_A) ) (\gamma_B (1-\mathcal{P}_B) )\rho_{00}(t_0) +\nonumber \\
             & &(\gamma_A (1-\mathcal{P}_A) ) (1 - \gamma_B \mathcal{P}_B ) \rho_{11}(t_0) +\nonumber \\
             & &(1-\gamma_A \mathcal{P}_A ) (\gamma_B (1 - \mathcal{P}_B) ) \rho_{22}(t_0) + \label{eksol10} \\
             & &(1- \gamma_A \mathcal{P}_A ) (1 - \gamma_B \mathcal{P}_B ) \rho_{33}(t_0). \nonumber
\end{eqnarray}

Comparing (\ref{Rho01})-(\ref{Rho33}) and (\ref{eksol1})-(\ref{eksol10}), one can find the conditions for
which the combined action of the three quantum channels is
equivalent to the Redfield theory for the quadrupolar relaxation: $\gamma_A  = 1-e^{-2CJ_{2}t}$,
 $\gamma_B  = 1-e^{-2CJ_{1}t}$, $\lambda = \frac{1}{2}\left(1+e^{-CJ_{0}t}\right)$
and $\mathcal{P}_A = \mathcal{P}_B = 1/2$. Note that the last condition implies that the
equilibrium state is a maximally mixed state, which corresponds to
the equilibrium state at infinite temperature. Nuclear spins under 
Zeeman interaction at room temperature correspond to a low polarized system, ($
\mathcal{P} \approx 1/2$ for all spins). Thus this requirement is
quite reasonable, as we will show comparing the theoretical
prediction to the experimental data in the next section. Furthermore, it is important
to emphasize that a high temperature approximation is implied in the solutions of 
Redfield equations (\ref{Rho01})-(\ref{Rho33}).

Therefore, a quantum circuit model for the
quadrupolar relaxation can be constructed combining the quantum circuits (\ref{gad3}) and
(\ref{gpd}). The circuit equivalent to the relaxation is shown in Figure (\ref{circ7}) and
 involves 7 qubits. The first two qubits, from top to bottom, correspond to the qubits
of the quadrupolar spin. The remaining five qubits correspond to
the environment, which is initialized in the state $| 000 \rangle
\otimes | \Phi \rangle_{eq}$, where  $| \Phi \rangle_{eq} =
(\sqrt{\mathcal{P}_A}| 0 \rangle + \sqrt{1-\mathcal{P_A}}| 1
\rangle) \otimes (\sqrt{\mathcal{P}_B}| 0 \rangle +
\sqrt{1-\mathcal{P}_B}| 1 \rangle$). The relations between the angles
$\theta$, $\alpha_{A,B}$, $\beta_{A,B}$ and the spectral densities
are summarized in the table (\ref{tabr}).

\begin{table}[htbp]
\tcaption{\label{tabr} Relation between $\theta$, $\alpha_{A,B}$, $\beta_{A,B}$ and the spectral densities.}
\begin{tabular}{c}\\
\hline
$\alpha_A = 2\arcsin\left( \sqrt{1-e^{-2CJ_{2}t}} \right )$ \\ \\
$\alpha_B = 2\arcsin\left( \sqrt{1-e^{-2CJ_{1}t}} \right )$ \\ \\
$\beta_A = 2\arccos\left( \sqrt{1-e^{-2CJ_{2}t}} \right )$ \\ \\
$\beta_B = 2\arccos\left( \sqrt{1-e^{-2CJ_{1}t}} \right )$ \\ \\
$\theta  = 2\arccos\left( 2 e^{-CJ_{0}t} \right )$ \\
\hline\\
\end{tabular}
\end{table}

\begin{figure}[htbp]
\vspace*{13pt}
\begin{center}

\begin{tabular}{ c }
\Qcircuit @C=2.3em @R=1.7em @!R {            & &  \ctrl{1} & \ctrlo{1}& \qw  \\        
                                             & & \ctrlo{1} &  \ctrl{1}& \qw \\
                                | 0 \rangle  & & \gate{R_y(\theta)} &  \gate{R_y(\theta)}& \qw  \\}

\end{tabular}

\end{center}
\vspace*{13pt}
\fcaption{\label{gpd} Quantum circuit description of the global phase damping channel. The first
two qubits (on the top of figure) correspond to the qubits of
the quadrupolar spin. The environment is comprised by only one qubit initialized in
the pure state $|0\rangle$. The notation $R_y(\theta)$ represents
a $\theta = 2\arccos(2\lambda - 1)$ rotation around the axis $y$.}
\end{figure}

\section{Experiment \label{expd}}
\noindent

The experiments were carried out using a 9.4 T -
VARIAN INOVA spectrometer on $^{23}$Na nuclei in a sample of a
lyotropic liquid crystal system (Sodium Dodecyl Sulfate = SDS) \cite{jcp2}.
Several pseudo-pure states were prepared using numerically
optimized pulses known as Strongly Modulated Pulses
(SMP)\cite{fortunato,suter2} and the evolution of each 
density matrix element was  
monitored by quantum
state tomography \cite{pra,jmr,teles}. More
specifically, it was experimentally followed the time evolution of
the pseudo-pure states \cite{oliveira}:

\noindent
\begin{equation}
\rho _{pps}=\frac{(1-\epsilon )}{2^{N}}\hat{\emph{1}}+\epsilon |\psi \rangle
\langle \psi |  \label{pps},
\end{equation}%
where $\epsilon \sim 10^{-6}$ is the nuclear polarization  at room
temperature and $|\psi \rangle$ corresponds to $|00\rangle$,
$|01\rangle$, $|10\rangle$, $|11\rangle $
$\frac{1}{2}(|00\rangle+|01\rangle+|10\rangle+|11\rangle)$, and
$\frac{1}{\sqrt{2}}(|00\rangle+|11\rangle)$.

\begin{figure}[htbp]
\vspace*{13pt}
\includegraphics[width=8.0cm]{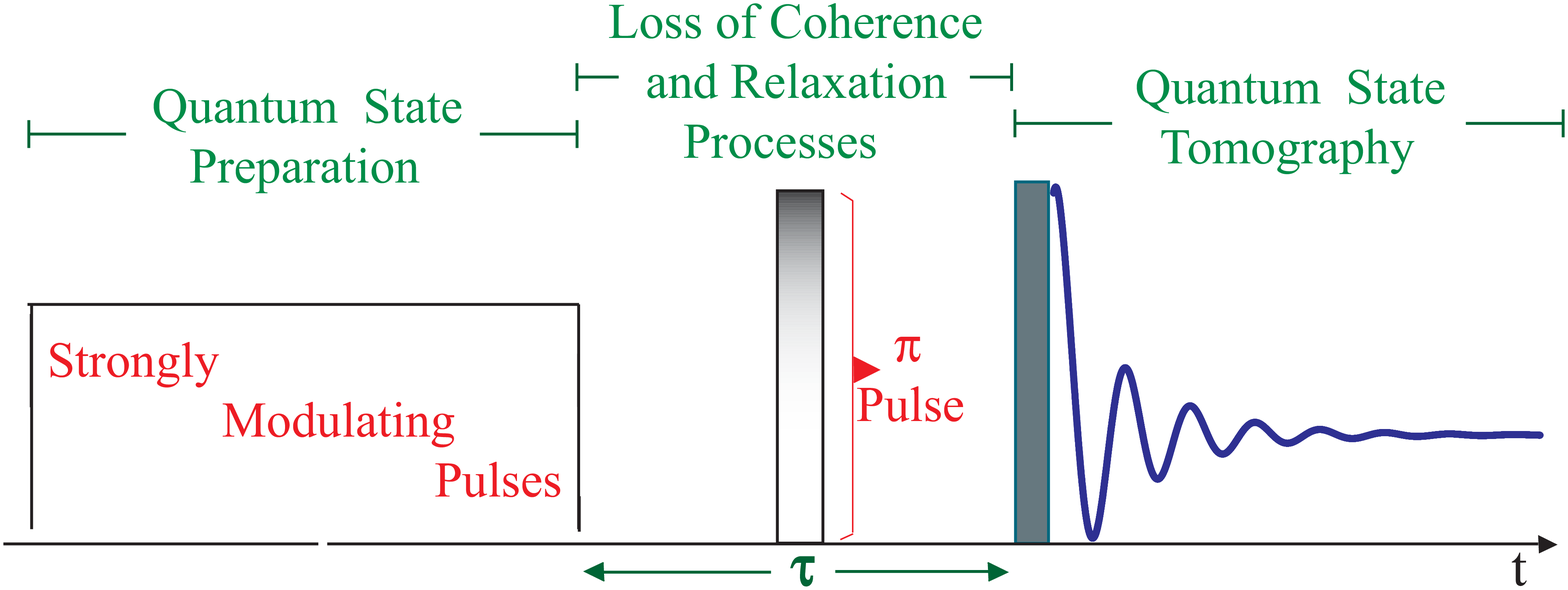}
\vspace*{13pt}
\fcaption{\label{SMP} Scheme of the pulse sequence used for
probing the relaxation of the individual density matrix elements.
The initial states are prepared with the SMP technique
\cite{fortunato}. The state relaxation takes place during a
variable evolution period and, finally, a hard RF pulse with the
correct phase cycling and duration is applied to execute quantum state tomography via
coherence selection \cite{teles}.}
\end{figure}

The basic experimental scheme presented in Figure (\ref{SMP})
consists of: a state preparation period performed using SMP
technique \cite{fortunato,suter2}; a variable evolution
period where relaxation processes
take place; a hard RF pulse with the correct phase cycling and
duration to execute quantum state tomography via coherence
selection \cite{teles}. For off-diagonal elements, a $\pi $
pulse was added in the middle of the evolution period to refocus
the $B_{0}$ field inhomogeneities. Since the quadrupolar evolution is
not refocused by the $\pi $ pulse, evolution periods multiple of
$2\pi /\omega _{Q}$ were used. The spectral densities were
experimentally determined in reference \cite{ruben} as being  $J_0 =
(14 \pm 1) \times 10^{-9}s$, $J_1 = (3.4 \pm 0.4) \times
10^{-9}s$, and  $J_2 = (3.7 \pm 0.3) \times 10^{-9}s$ and $C =
(1.2 \pm 0.1) \times 10^{10} s^{-2}$.

Using the experimental spectral densities and the initial
density matrix as inputs, we have simulated the
evolution of density matrix (\ref{pps}) using the
7-qubit circuit model shown in (\ref{circ7}) and then compared
our simulation with experimental data. The comparisons
for populations of deviation density matrix ($\Delta \rho =  \rho_{pps} - I/4$ \cite{oliveira}) corresponding to 
the states $\left|01\right\rangle$,
$\left|10\right\rangle$, $\left|11\right\rangle$,
$\frac{1}{\sqrt{2}}(|00\rangle+|11\rangle)$ and
$\frac{1}{2}(|00\rangle+|01\rangle+|10\rangle+|11\rangle)$ are
shown in Figure \ref{figpol}. The results for the non-diagonal
elements corresponding to an uniform
superposition are presented in Figure (\ref{figsup}). In all cases,
the experimental behavior could be well reproduced using the
circuit model (\ref{circ7}). The good agreement between the
theoretical and the experimental results shows that the electric quadrupolar  nuclear
relaxation can be viewed as three different
channels and that the circuit derived in this work provides a good quantum computation
model for the relaxation. Small discrepancies between the experimental and theoretical 
results are due to experimental errors in the determination of the spectral densities 
and initial density matrices.

\begin{figure*}[htbp]
\vspace*{13pt}
\begin{center}

\begin{tabular}{ c }
\Qcircuit @C=0.4em @R=0.1em @!R { && & \ctrl{1}          & \ctrlo{1}         & \ctrl{3}            &   \targ  & 
\qw      & \ctrlo{3}& \targ & \gate{Z}&\qswap & \qw&\qw&\qw& \qw& \qw&\qw&\qw&\qw&\\
                                     && & \ctrlo{1}         & \ctrl{1}          &  \qw                &     \qw  & 
\qw      & \qw& \qw&\qw&\qw \qwx& \ctrl{3}&\targ&\qw      & \ctrlo{3}& \targ & \gate{Z}&\qswap &\qw&\\                    
                      | 0 \rangle    && & \gate{R_y(\theta)}& \gate{R_y(\theta)}&  \qw                &    \qw   &
\qw      & \qw&\qw&\qw&\qw \qwx&\qw&\qw&\qw& \qw& \qw&\qw&\qw\qwx&\qw&\\
                      | 0 \rangle    && & \qw               & \qw               & \gate{R_y(\alpha_A)}& \ctrl{-3}&
\targ    & \gate{R_y(\beta_A)}& \ctrlo{-3}& \ctrl{-3}&\qswap \qwx& \qw&\qw&\qw& \qw& \qw&\qw&\qw \qwx&\qw&\\
                      | 0 \rangle    && & \qw               & \qw               &      \qw            &    \qw   &
\qw      & \qw&\qw&\qw&\qw&\gate{R_y(\alpha_B)}&\ctrl{-3}&\targ    & \gate{R_y(\beta_B)}& \ctrlo{-3}& \ctrl{-3}&\qswap \qwx&\qw&\\
                            && & \qw               & \qw               &     \ctrlo{-2}      &\ctrlo{-2}&
\ctrl{-2}&\ctrl{-2}&\ctrl{-2}&\ctrl{-2}&\ctrl{-2}&\qw&\qw&\qw& \qw& \qw&\qw&\qw&\qw&\\
                                     && & \qw               & \qw               &      \qw            &     \qw  &
\qw      & \qw&\qw&\qw&\qw&\ctrlo{-2}&\ctrlo{-2}&\ctrl{-2}&\ctrl{-2}&\ctrl{-2}&\ctrl{-2}&\ctrl{-2}&\qw&\\}

\end{tabular}

\end{center}
\vspace*{13pt}
\fcaption{\label{circ7}  Quantum circuit description of the nuclear spin $3/2$ electric
quadrupole relaxation. The first two qubits (on the top of figure) correspond to the qubits of
the quadrupolar spin. The remaining five qubits corresponds to the environment which is
initialized in the state $| 000 \rangle \otimes | \Phi \rangle_{eq}$
where  $| \Phi \rangle_{eq}  = (\sqrt{\mathcal{P}_A}| 0 \rangle + \sqrt{1-\mathcal{P_A}}| 1 \rangle) \otimes.
(\sqrt{\mathcal{P}_B}| 0 \rangle + \sqrt{1-\mathcal{P}_B}| 1 \rangle$).}
\end{figure*}

\section{Conclusions}
\noindent
In this work, we have studied the relaxation of a
nuclear quadrupolar system, originated by local electric field gradient
fluctuations, using  the language of quantum circuits and following the approach of quantum
information processing. We were able to
describe the quadrupolar relaxation by a model in which the environment is comprised by five qubits and
three different quantum noise channels acting on the
quadrupolar system. The interaction between the environment and a spin 3/2 was described by a
quantum circuit fully compatible with the Redfield theory of relaxation. The theoretical predictions
 were compared to experimental results. The good agreement between the
theoretical and the experimental results shows that the model derived in
this works provides a good quantum computation
model for the relaxation.

Although many studies concerning open systems have been carried out
for decades, just few studies consider relaxation phenomena
in the context of information processing. We believe that further efforts on
this direction could bring a better understanding about open systems
as well to yield some insight into the problem of simulating open
systems on quantum computers.

\begin{figure*}[htbp]
\vspace*{13pt}
\begin{center}
\subfigure[$| \psi \rangle = \left|00\right\rangle$ ] {\includegraphics[width=6.0cm]{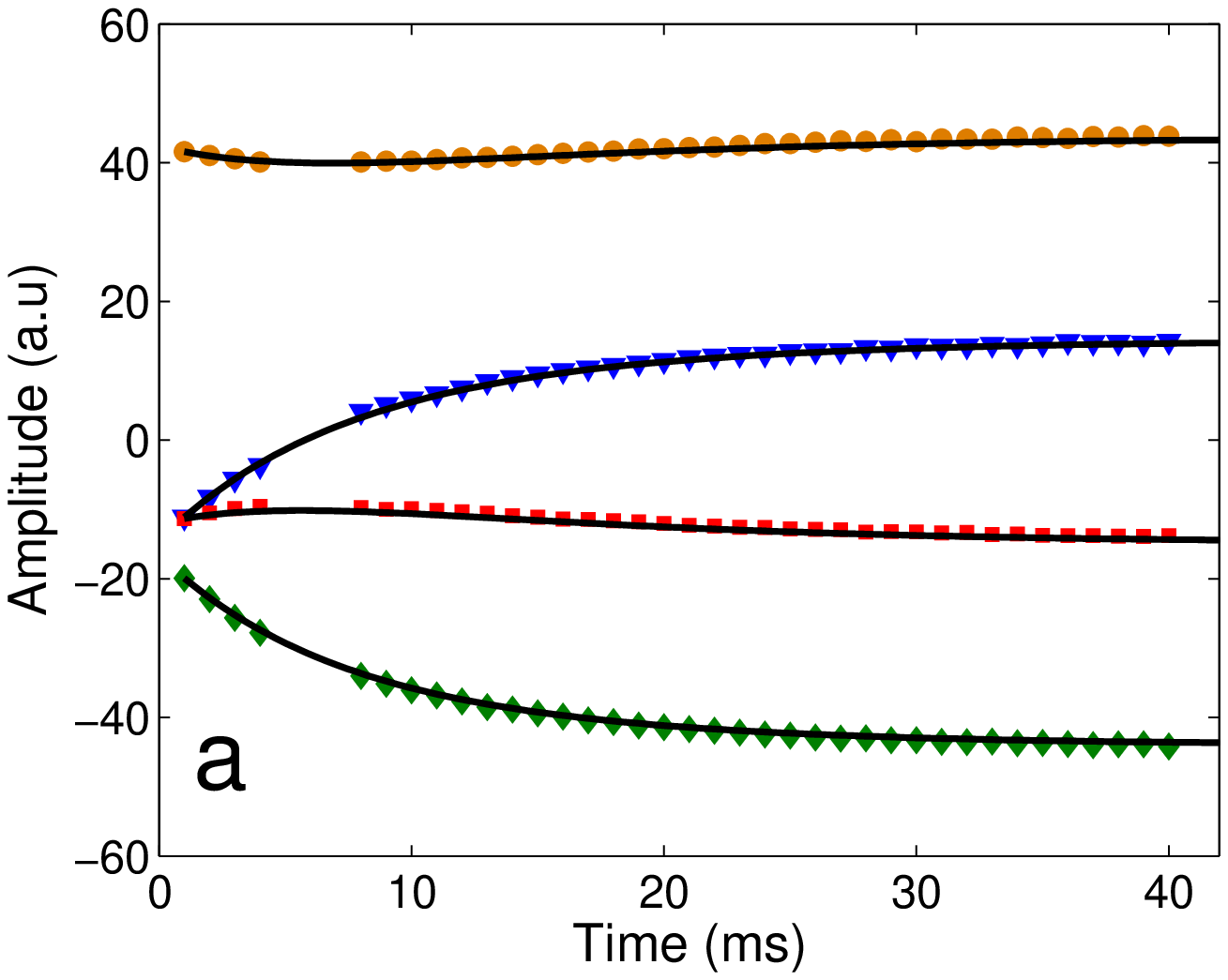}}
\subfigure[$| \psi \rangle = \left|01\right\rangle$ ] {\includegraphics[width=6.0cm]{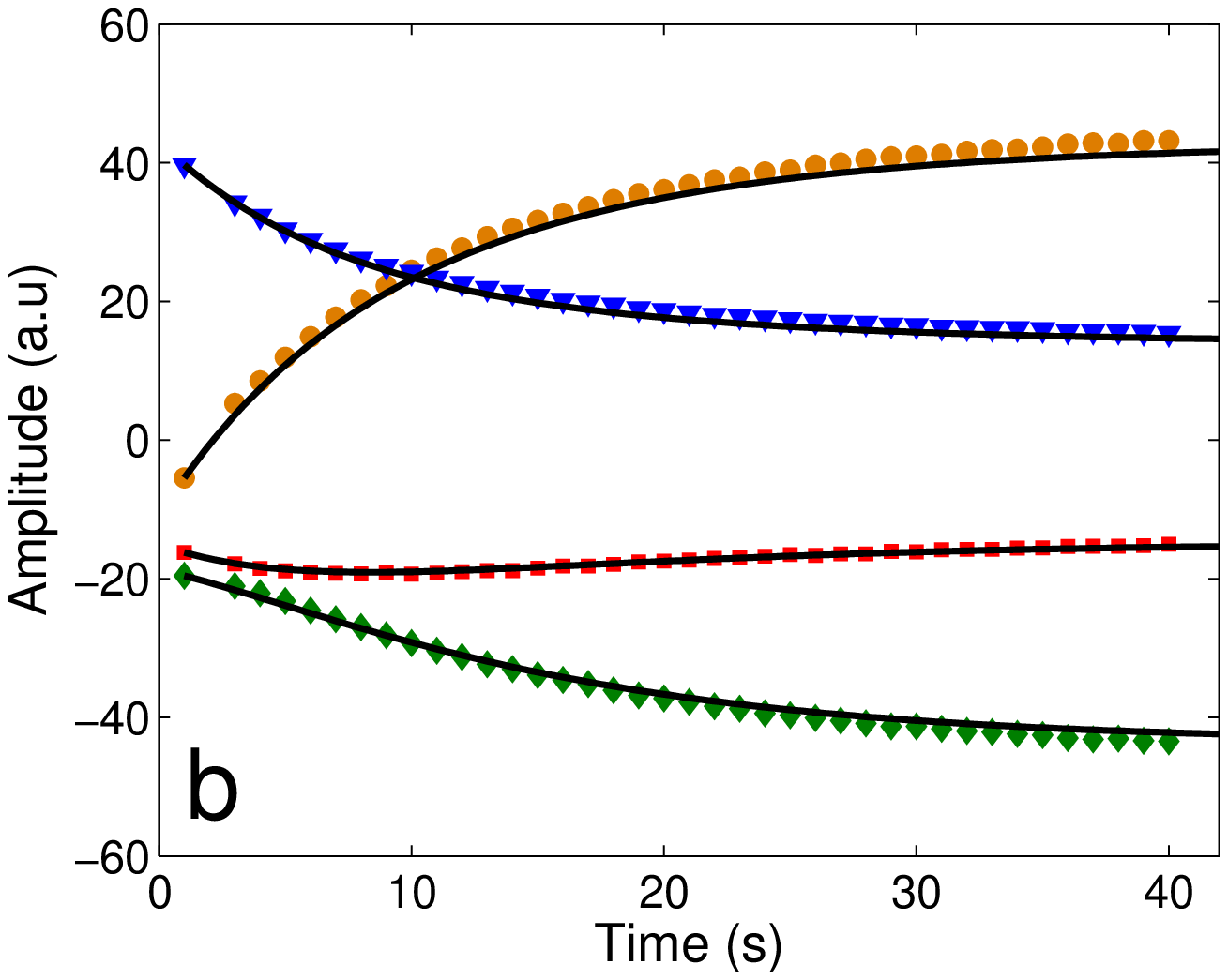}}\\
\subfigure[$| \psi \rangle = \left|10\right\rangle$ ] {\includegraphics[width=6cm]{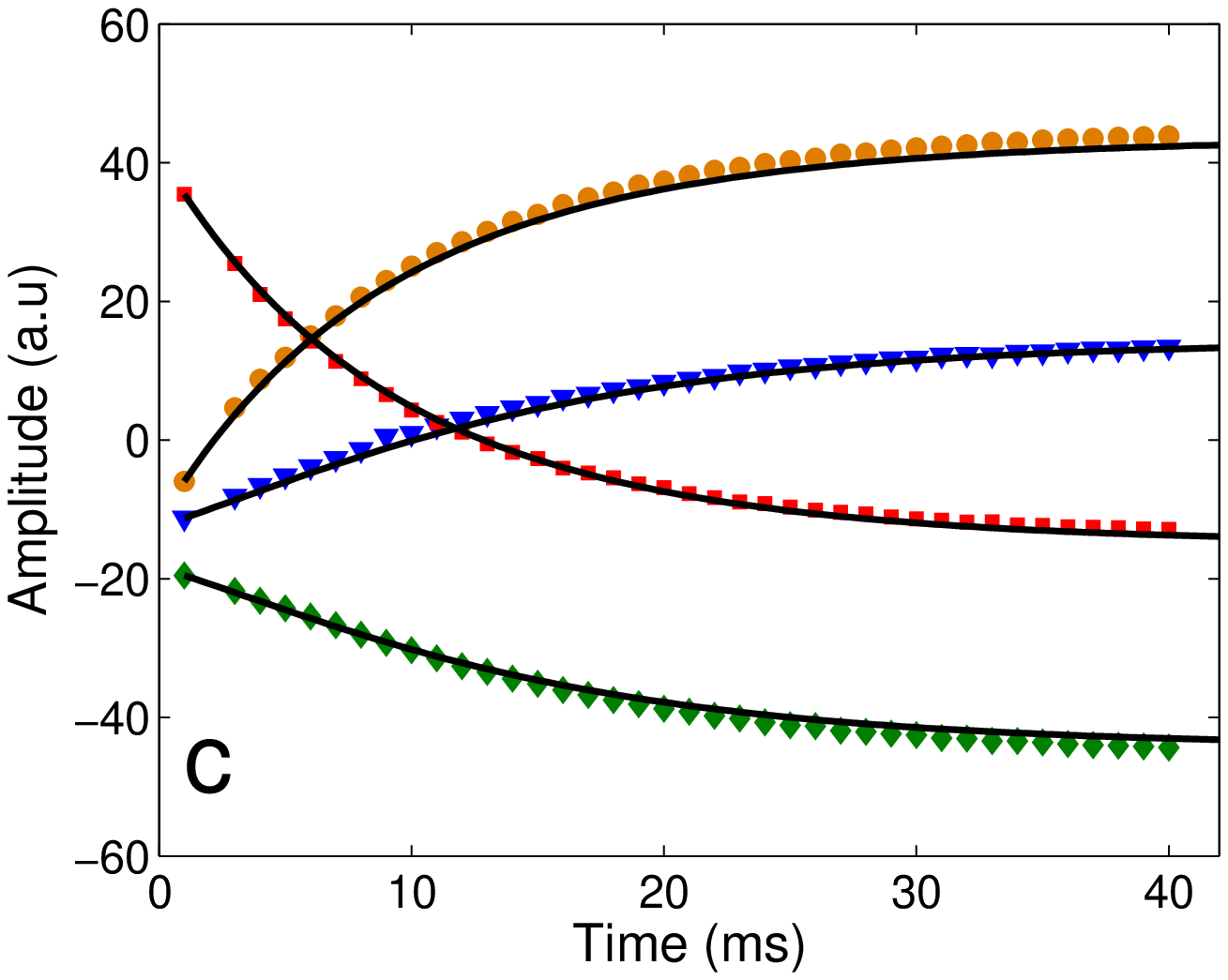}}
\subfigure[$| \psi \rangle = \left|11\right\rangle$ ] {\includegraphics[width=6cm]{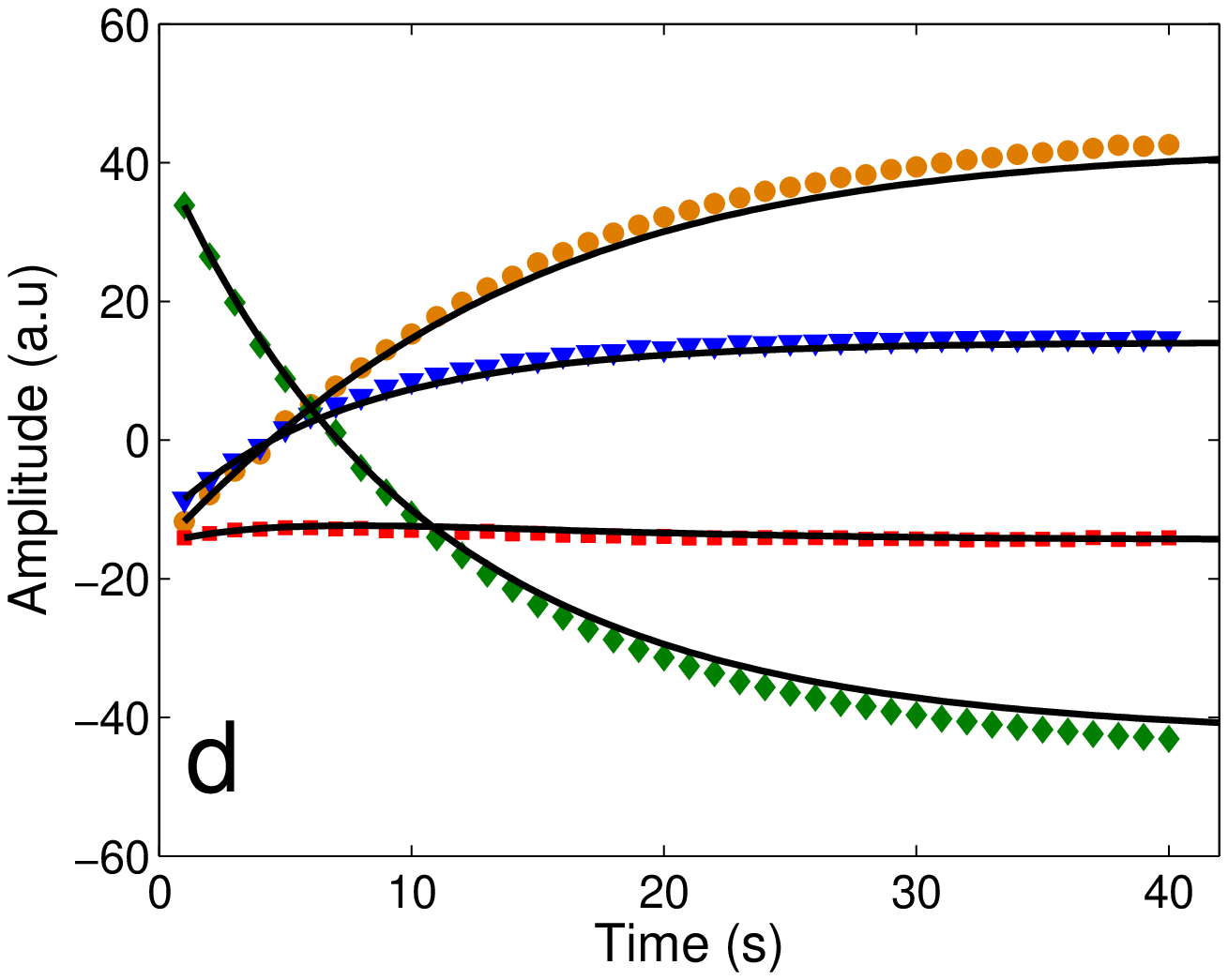}}\\
\subfigure[$| \psi \rangle = \frac{1}{\sqrt{2}}(|00\rangle+|11\rangle)$ ] {\includegraphics[width=8cm]{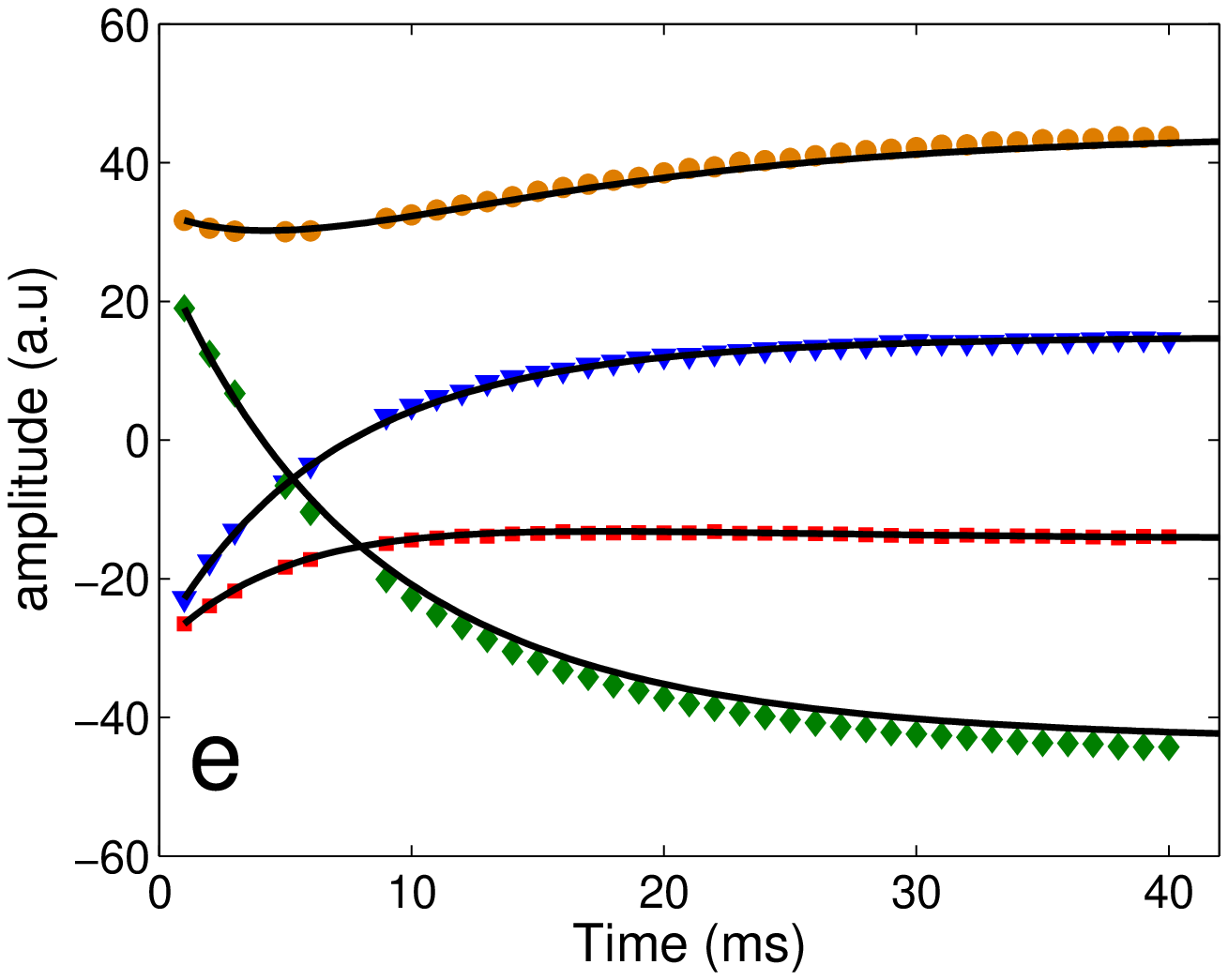}}
\end{center}
\vspace*{13pt}
\fcaption{\label{figpol} Combined experimental data ( {\color{yellow}{$\bullet$}}-$\rho_{00}$,
{\color{blue}{$\blacktriangledown$}}-$\rho_{11}$,  {\color{red}{$\blacksquare$}}-$\rho_{22}$,
{\color{green}{$\Diamond$}}-$\rho_{33}$  ) and the corresponding theoretical
prediction (solid lines) obtained from the circuit (\ref{circ7}) for various 
input states. The experimental data were obtained  
from previous experiments \cite{ruben}.}
\end{figure*}

\begin{figure*}[htbp]
\vspace*{13pt}
\begin{center}
\subfigure[{\color{yellow}{$\bullet$}}-$\rho_{00}$,
{\color{blue}{$\blacktriangledown$}}-$\rho_{11}$, {\color{red}{$\blacksquare$}}-$\rho_{22}$,
{\color{green}{$\Diamond$}}-$\rho_{33}$.] {\includegraphics[width=6.0cm]{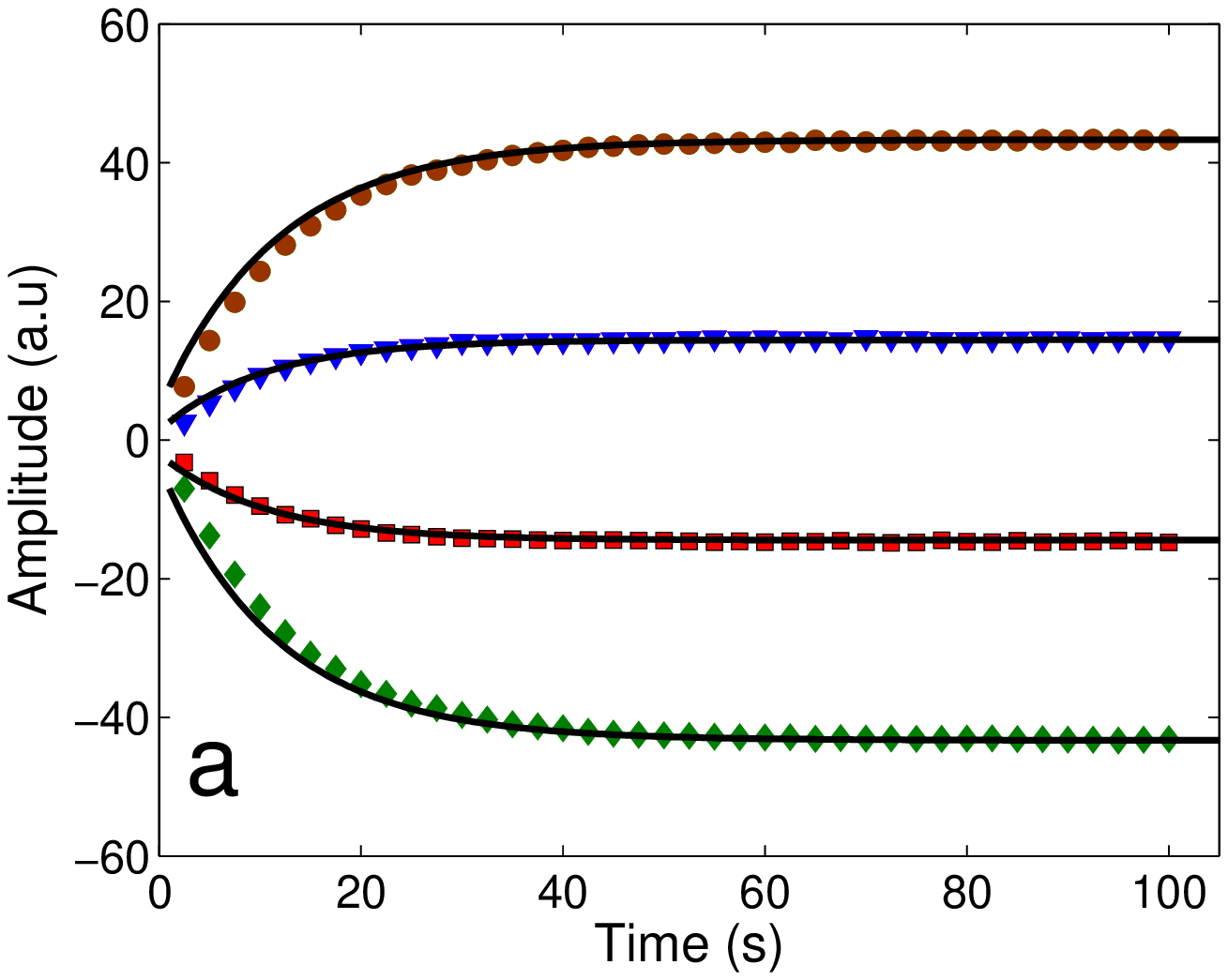}}
\subfigure[{\color{blue}{$\bullet$}}-$\rho_{02}$,
{\color{red}{$\blacktriangledown$}}-$\rho_{13}$] {\includegraphics[width=6.0cm]{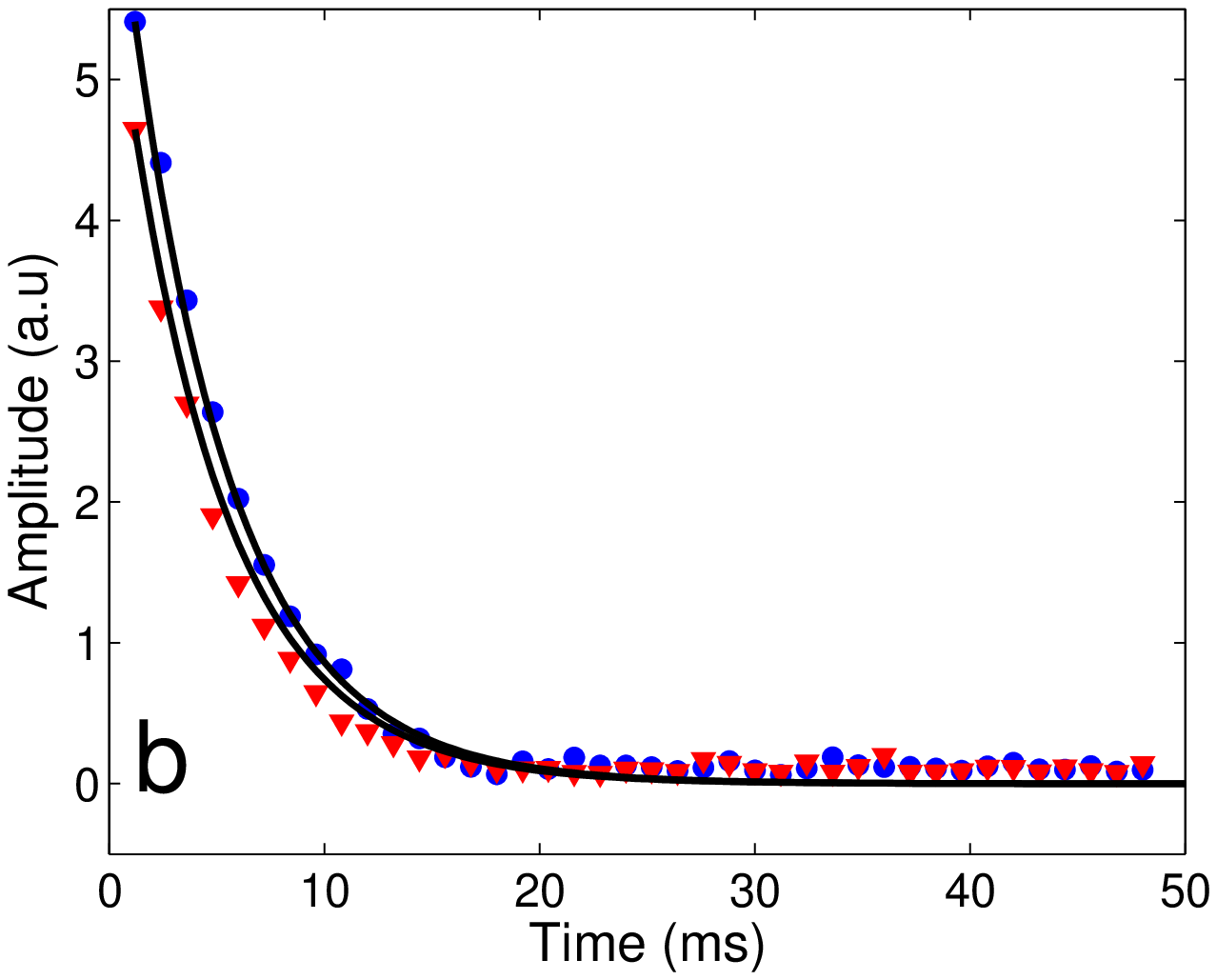}} \\
\subfigure[{\color{blue}{$\bullet$}}-$\rho_{12}$,
{\color{red}{$\blacktriangledown$}}-$\rho_{03}$] {\includegraphics[width=6cm]{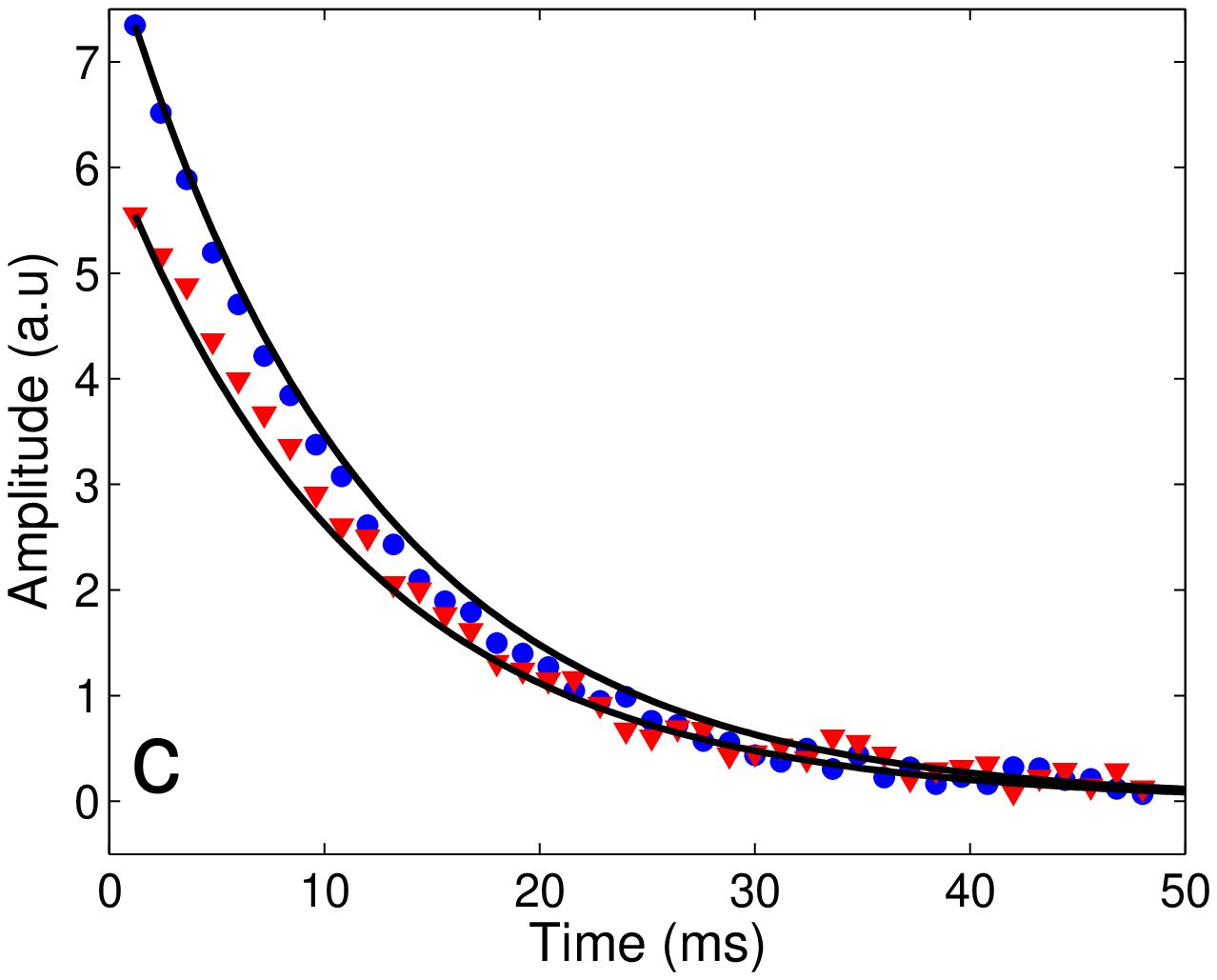}}
\end{center}
\vspace*{13pt}
\fcaption{ \label{figsup} Combined experimental data (symbols)
and the corresponding theoretical prediction (solid lines) obtained from
the circuit (\ref{circ7}) and  starting from the
state $\frac{1}{2}(|00\rangle+|01\rangle+|10\rangle+|11\rangle)$. The experimental data were obtained  
from previous experiments \cite{ruben}. \label{exp2}}
\end{figure*}

\nonumsection{Acknowledgements}
\noindent
The authors acknowledge the financial support of the Brazilian
Science Foundations CAPES, CNPq and FAPESP. We also thank the
support of the Brazilian network project National Institute for
Quantum Information. AMS would like to acknowledge the government 
of Ontario - Canada.

\nonumsection{References}

\end{document}